\documentclass{pasj00} 
\begin{document}
\SetRunningHead{Y. Sofue, et al.}{Galactic Constant Determination}
\Received{2010/mm/dd}  \Accepted{2010/mm/dd} 

\def\b{\bf}
\def\kms{km s$^{-1}$}  \def\Msun{M_\odot} 
\def\be{\begin{equation}} \def\ee{\end{equation}}
\def\bc{\begin{center}} \def\ec{\end{center}}
\def\Rsun{R_0} \def\Vsun{V_0} \def\sin{{\rm ~sin~}} \def\cos{{\rm ~cos~}}
\def\vr{v_{\rm r}} \def\vp{v_{\rm p}}  \def\p{$\pm$}
\def\sinl{~{\rm sin~}l~} \def\cosl{~{\rm cos~}l~}
\def\kag{Dept. Physics and Astronomy, Kagoshima-University, Kagoshima 890-0063}
\def\mei{Dept. Physics, Meisei University, Hino, Tokyo 191-8506}
\def\nao{Mizusawa VLBI Observatory, National Astronomical Observatory of Japan, Mitaka, Tokyo 181-8588}
\def\ioa{Inst. Astronomy, Univ. Tokyo, Mitaka 181-0015, Tokyo}

\title{Near-Solar-Circle Method for Determination of the Galactic Constants}
\author{Sofue, Y.$^{1}$, Nagayama, T.$^{2}$, Matsui, M.$^3$ , and Nakagawa, A.$^3$ }  
\affil{1. \mei   \\ 
\& \ioa, \\
2. \nao, \\
3. \kag, \\ 
E-mail: sofue@ioa.s.u-tokyo.ac.jp}


\maketitle

\begin{abstract}   

We propose a method to determine the galactic constants $R_0$ (distance to the Galactic Center) and $V_0$ (rotation velocity of the Sun) from measurements of distances, radial velocities and proper motions of objects near the solar circle. This is a modification of the solar-circle method to a more practical observational method. We apply the method to determine $R_0$ using data from the literature with known distances and radial velocities, and obtain  $R_0=7.54 \pm 0.77$ kpc. 
\end{abstract}

\section{Introduction}
 
The major parameters for studying the structure and dynamics of the Galaxy are the galactic constants $R_0$ and $V_0$, which are the distance of the Sun from the Galactic Center and the circular rotational velocity of the LSR (Local Standard of Rest), respectively. They are the most fundamental parameters for the rotation curve and mass analyses of the Galaxy, and are often assumed to be 8 kpc and 200 \kms (Sofue et al. 2009). However, the currently determined values allow a wide range of uncertainties, diverging from $\sim 7$ to 9 kpc and $\sim 180$ to 250 \kms (Reid et al. 1993, 2009a; McMillan and Binney 2010; Olling and Merrifield 1998;  Honma and Sofue 1997). 

There have been various methods to  determine $R_0$ and $V_0$, that include the direct distance measurements of the parallax of Sgr A$^*$ using radio VLBI technique, measurement of the distance to the star forming region Sgr B using the statistical parallax method, and the measurement of distance to the center of mass of the distribution of globular clusters from spectroscopic parallax as well as the period-luminosity relation of RR Lyr variables (review by Reid et al. 1993;  Reid et al. 1988, 2004, 2009a, b; Eisenhauer et al. 2005; McMillan and Binney 2010). 

The solar-circle method is a geometrical method to determine $R_0$, in which an object with known distance and zero LSR velocity is used to solve an isosceles triangle as illustrated in figure \ref{fig-circ0} (Miharas 1981). It requires no other assumption, except for the circular rotation. The rotation velocity of the Sun $V_0$ is also determined, if the proper motion of the object is measured. However, since the source is required to lie exactly on the solar circle, the method has been rarely applied. Ando et al. (2011) recently applied this method to the star forming region ON2N, which lies exactly on the solar circle with zero LSR radial velocity.

In the present paper, we modify the solar-circle method, which requires sources with LSR radial velocity $\vr=0$ \kms, to  a more practical way, so that a larger number of galactic objects, e.g. with $|\vr|\le 15$ \kms, may be used.  We present formulation of the method and estimates. The method makes it possible to directly estimate the galactic constants without assuming a rotation curve. In this context, it may be complimentary to the statistical likelihood method  assuming a rotation curve engaged by McMillan and Binney (2011). We also try to apply the method to determine $R_0$ using data from the literature for HII regions with known distances.  

\section{The Near-Solar-Circle Method}

If an object is located exactly on the solar circle, its line of sight velocity is zero, $\vr=0$, and the galactic constants, $\Rsun$ and $\Vsun$, are determined simply by measuring the distance, $r$, and perpendicular velocity in the direction of galactic longitude, $\vp$ as
\be
R_0 ={r \over 2 \cos l}, \label{eqr00}
\ee
and 
\be
V_0 =-{\vp \over 2\cos l}. \label{eqv00}
\ee
  Here, and hereafter, the velocities $\vr$ and $\vp$ are referred to the LSR coordinates after correction for the solar motion. 

It is, however, seldom to find an object exactly located on the solar circle with the radial velocity being equal to zero. We therefore consider using objects which are near the solar circle with finite $\vr$, and modify the solar circle method for more practical observations. 

\begin{figure}  
\FigureFile(80mm,80mm){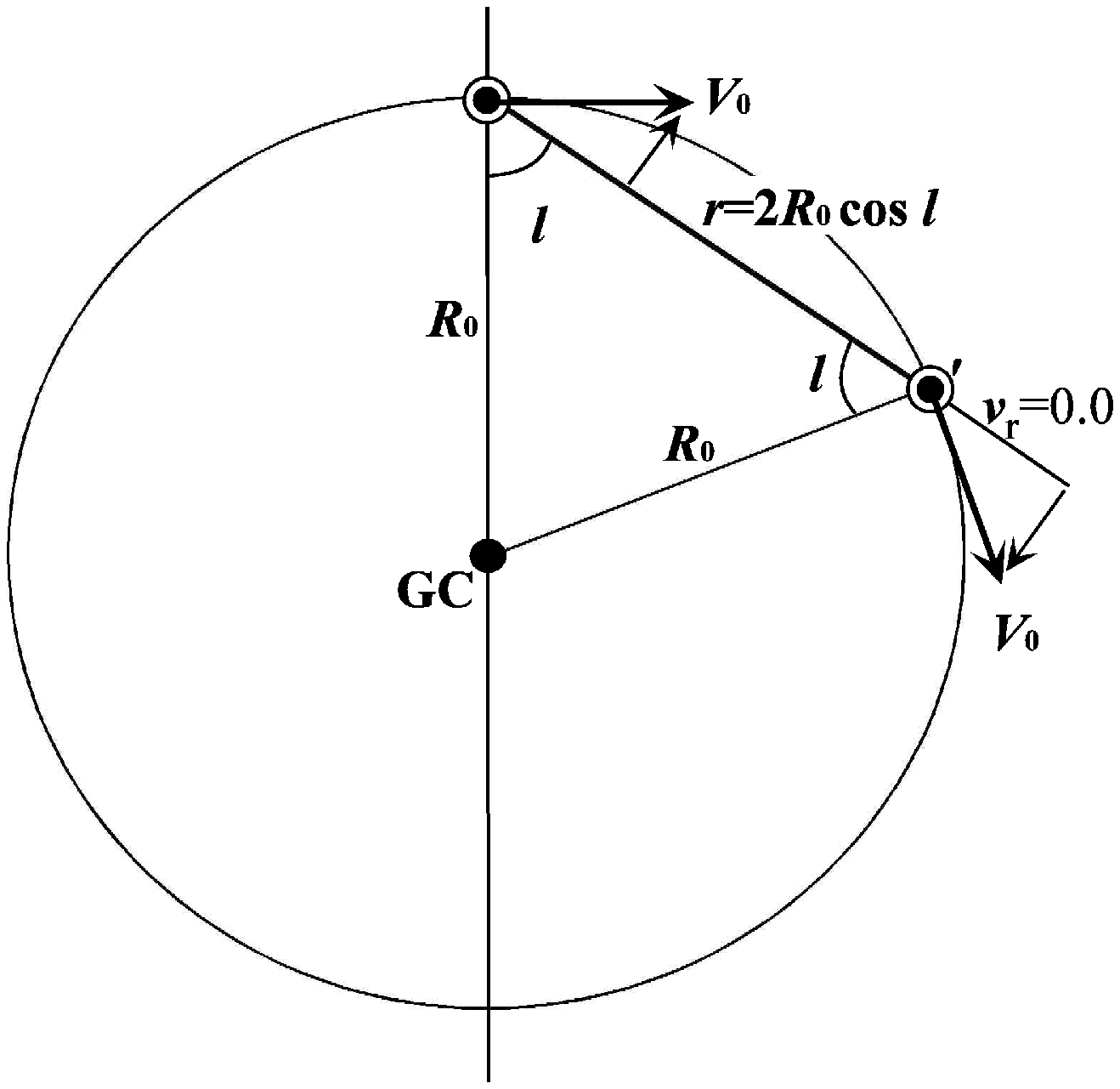}
\caption{Solar-Circle method to determine $R_0$, using an object with $\vr=0.0$ \kms.}
\label{fig-circ0}   

\FigureFile(80mm,80mm){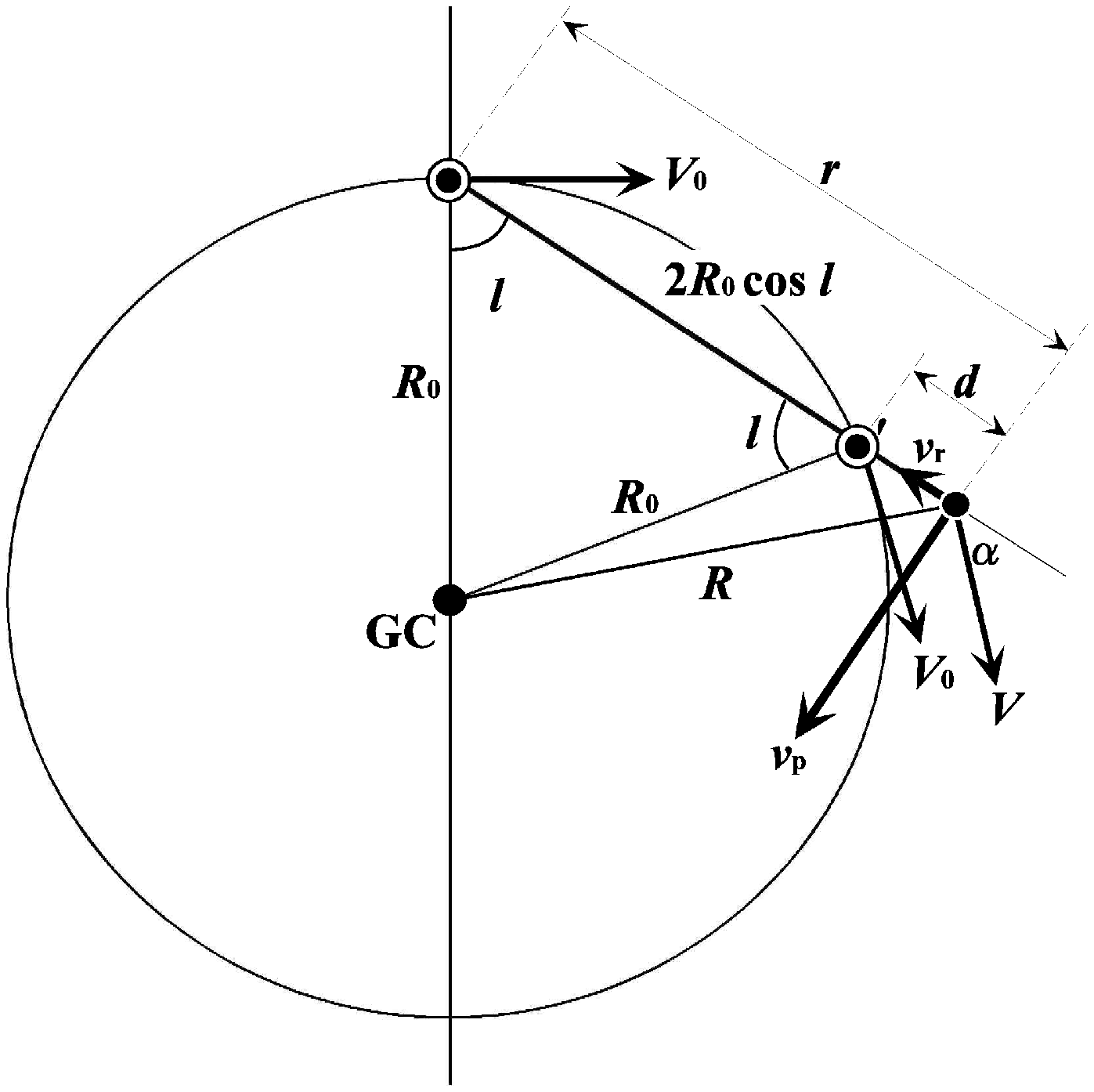}
\caption{Near-solar-circle method to determine the Galactic constants $\Rsun$ and $\Vsun$.}
\label{fig-circ}  
 \end{figure}

We consider an object in the galactic plane at galactic longitude $-90^\circ<l<90^\circ$ at a distance from the Sun $r$. We denote the distance of the object from the solar circle on the line of sight by $d$ as illustrated in figure \ref{fig-circ}. Then, $r$ and $d$ are related to the galacto-centric distance $R$ and longitude $l$ by
\be
 r = 2 \Rsun \cos l +d.
\label{eq-r}
\ee

The radial velocity, $\vr$, of the object is expressed by
\begin{equation}
 \vr = \left( V{\Rsun \over R} - \Vsun \right) \sin~ l,
\end{equation}
where $V$ is the rotation velocity of the object at a galacto-centric distance $R$. If the object is near the solar circle, $d\ll r$. this is written using $d$ as
\begin{equation}
 \vr = - A~d~\sin~ 2l.
\label{eq-vr}
\end{equation}
Note that $\vr$ is negative for an object outside the solar circle as shown in figure \ref{fig-circ}, for which $d$ is positive. 
Here, $A$ is the Oort constant,
\begin{equation}
 A = {1 \over 2} \left[ {\Vsun \over \Rsun} - {dV \over dR} \right]_{R=\Rsun}.
\label{eq-A}
\end{equation} 
 
The perpendicular component, or the proper motion, is written as
\begin{equation}
 \vp = {V \over R}(\Rsun  \cos~l -r) -\Vsun \cos~l.
\label{eq-vp}
\end{equation} 
This value is negative for object's moving toward decreasing longitude as for the case shown in figure \ref{fig-circ}.

We now consider a case that the object is not on the solar circle, as usually is the case, but it is located near the circle so that $|d| \ll r$ and $|v_{\rm r}|\ll \Vsun$, e.g. $|v_{\rm r}|< \sim 15$ \kms. We may Taylor expand equations \ref{eq-vr} and \ref{eq-vp} in terms of quantities including $d(\ll r)$ and $\vr$. We obtain the following explicit expressions of $R_0$ and $V_0$.
\be
\Rsun = {r\over 2\cos l}\left(1+{\vr\over A r \sin 2l}\right)
={r\over 2\cos l}\left(1-{d \over r}\right) 
\label{eq-rsun}
\ee
and
\be
\Vsun 
=-{\vp\over2\cos l}\left(1-{d\over r}-2{\vr\over\vp}{\cos^2l\over\sin l}\right)
\label{eq-vsun}
\ee
$$
=-{\vp\over2\cos l}\left(1-{d\over r}\right) + \vr {\rm cot}~ l. 
$$
Here, we neglected the second order terms of $d$ and $\vr$ which is small compared to $V_0$ or $|\vp|$. The above expressions include the Oort constant $A$, which includes $R_0$ and $V_0$. Since the term including $A$ is multiplied by $\vr$, the effect of the uncertainty of $A$ is of the second order magnitude. We adopt the most often used value of $A=15 \rm{km~s^{-1} {kpc}^{-1}}$. In this context, the present method is not perfectly independent of the current determination of the galactic constants. However, the effect from the uncertainty of $A$ is of the second order smallness, and the errors arising from an error of $A$ can be neglected in the present case.

\begin{figure}
\bc
\FigureFile(75mm,75mm){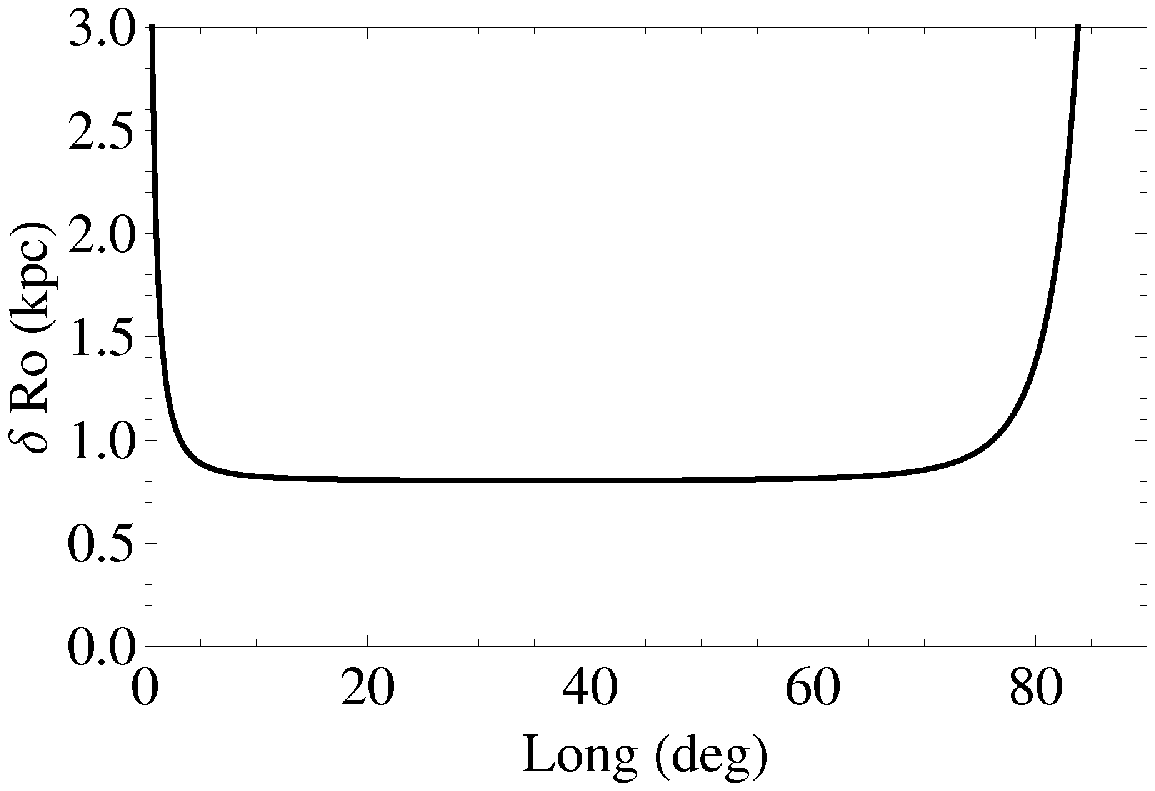}
\FigureFile(75mm,75mm){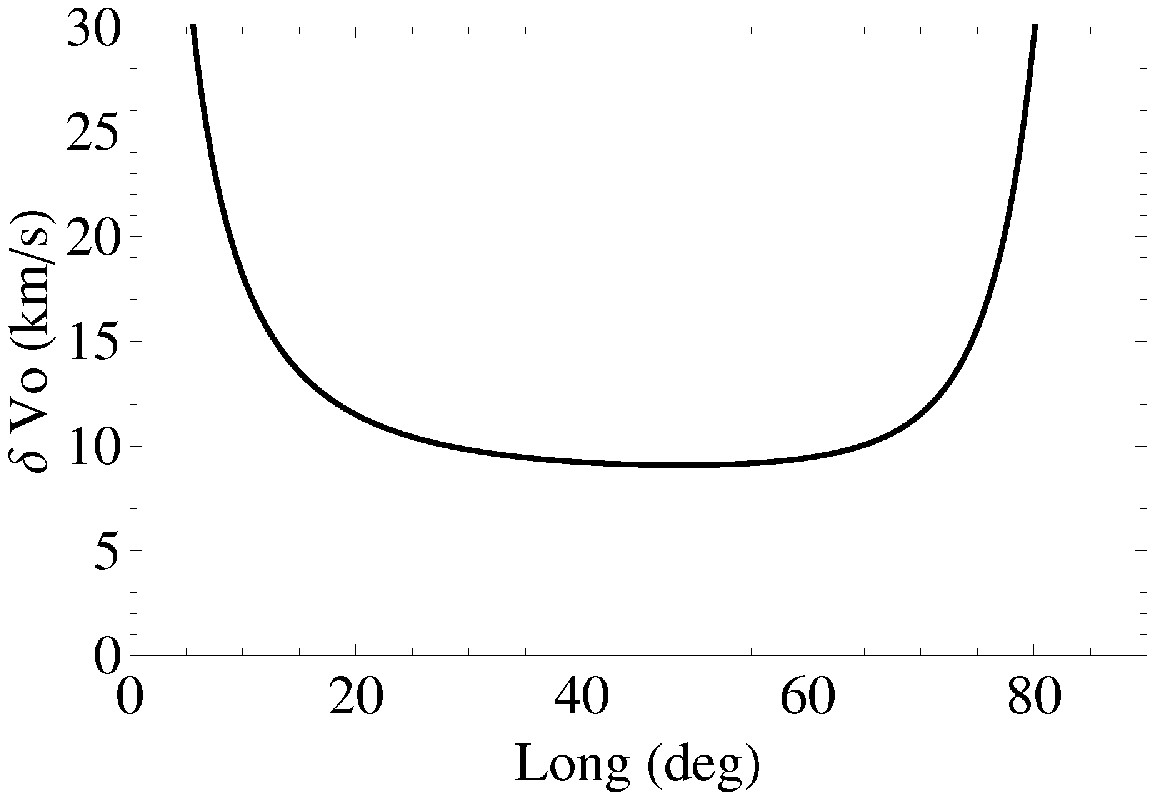}
\ec 
\caption{Error estimates for determination of $R_0$ and $V_0$ by using the near solar-circle method  plotted against   longitude $l$. For minimizing the errors both in $R_0$ and $V_0$, observations of objects near the solar circle at  $l=50-70^\circ$ are desirable. The errors of observables are fixed to be $\delta r=1$ kpc, and $\delta \vr=\delta \vp=10$ \kms.}
\label{fig-dR0V0}  
\end{figure}  

If we can measure the distance $r$, radial velocity $\vr$, and the proper motion $\vp$ of an object near the solar circle, e.g. with $|\vr|<15$ \kms, we may use  equations \ref{eq-rsun} and \ref{eq-vsun} to directly calculate the Galactic constants $\Rsun$ and $\Vsun$ at the same time. In order for the error to be small enough, the distance $r$ must be sufficiently large. Such observations have become recently possible indeed using VERA applying the VLBI technique (Honma et al 2007): We are able to determine $r$ by parallax measurements, $\vr$ by spectroscopy of maser radio line emissions, and $\vp$ by proper motion measurements.
\section{Error Propagation}

\def\dr{\delta r} \def\dvr{\delta \vr} \def\dvp{\delta \vp} \def\dA{\delta A}

The errors of $\Rsun$ and $\Vsun$ are caused by the observational errors $\dr$, $\dvr$, $\dvp$ of $r$, $\vr$, $\vp$, respectively, as well as by a possible error $\delta A$ of the adopted Oort constant $A$. We calculate the propagation of errors in equation \ref{eq-rsun} and \ref{eq-vsun}, but neglect the second order contributions including terms of the order of $(\delta x_i \delta x_j)^2$, where $\delta x_{i, j}=\dr, ~\vr, ~\dvr, ~\dvp ~{\rm or} ~\dA$. We obtain
\be
\delta \Rsun = {1 \over 2 \cos l}\left[\dr^2
+ \left(\dvr \over A \sin 2 l \right)^2 \right]^{1/2}
\label{eq-drsun}
\ee
and 
\be
\delta \Vsun = {1 \over 2 \cos l}
\left[\left({1 \over A r \sin 2 l}-{2 \cos^2 l \over \vp \sin l} \right)^2 
\vp^2\dvr^2 + \dvp^2\right]^{1/2}.
\label{eq-dvsun}
\ee
 
Here, the terms of smaller order than O($\vr \times \delta x^2$) with $x$ being $r$, $\vp$, $\vr$ or $A$, which are originally included in the parentheses, are neglected, because the radial velocity is small compared to the rotation velocities by the definition of the near-solar circle objects. 
Note also that the error $\delta A$ occurring from the uncertainty in $A$ appears in the second order terms, and has been neglected.

In figure \ref{fig-dR0V0} we show examples for the resulting errors in $R_0$ and $V_0$ calculated for a set of given values of errors of the observables as $\delta r = 1$ kpc, $\vr=10$ \kms and $\vp=10$ \kms.

\section{Application}

\subsection{Determination of $R_0$}

\begin{figure*}  
\bc
\FigureFile(120mm,120mm){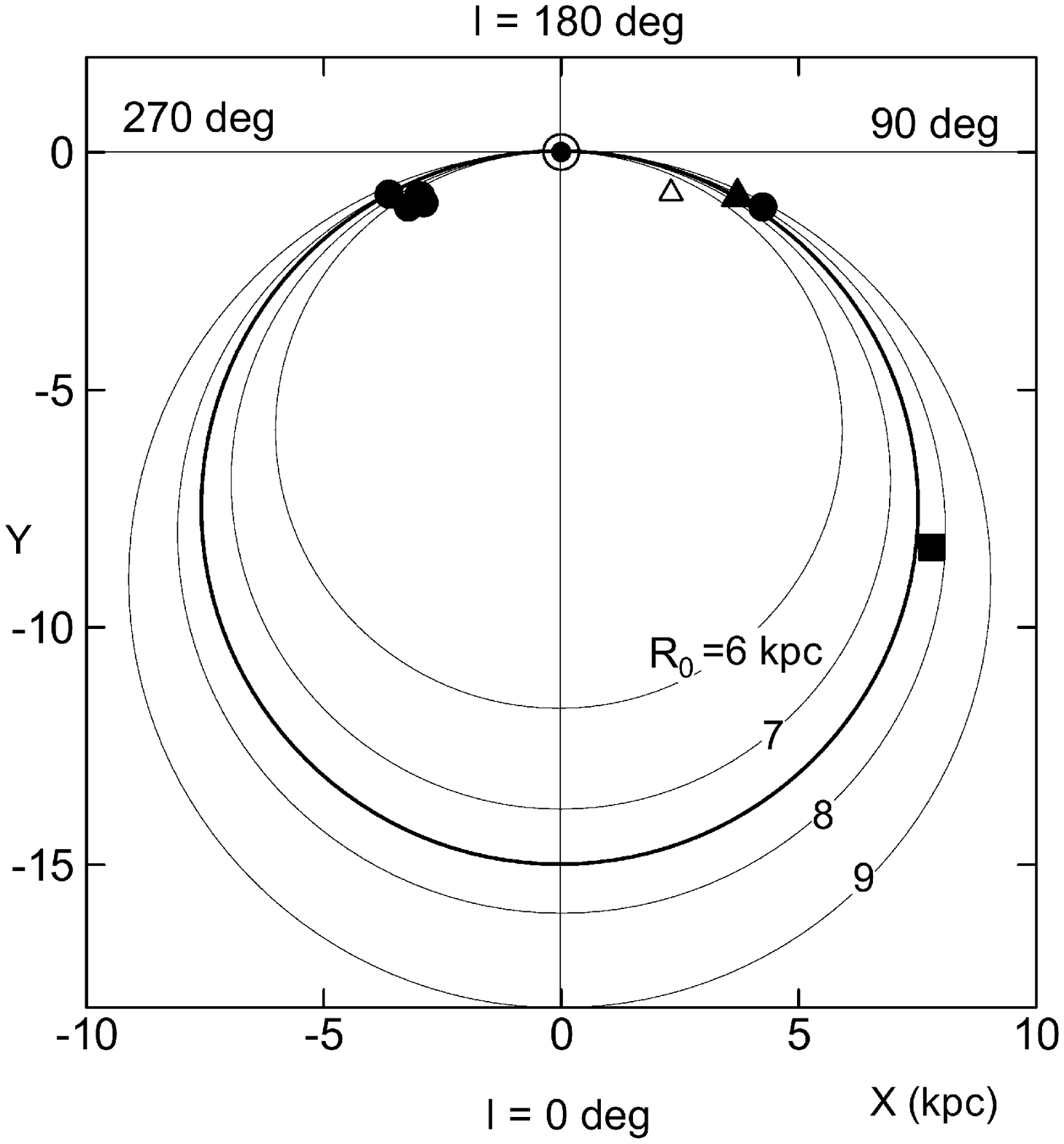}
\ec
\caption{  Distribution of the near-solar circle sources on the Galactic plane.  Filled circles are HII regions with photometric distances taken from Brand and Blitz (1993). Filled square is W49 from VLBI measurement by Gwinn et al (1992). Triangles are VERA sources with parallax distances and proper motions. Thin big circles denote possible solar circles with radii 6, 7, 8 and 9 kpc. The thick big circle is the solar circle for obtained value in this paper, $R_0=7.54 (\pm 0.77) $ kpc.} 
\label{plot} 
 \end{figure*}

Brand and Blitz (1993) compiled data sets of photometric (spectroscopic) distances of Galactic HII regions/reflection nebulae with measured radial velocities from associated molecular clouds. Their list includes a number of near-solar circle objects, which have sufficiently small radial velocities.  Considering the error estimates in figure \ref{fig-dR0V0}, we chose sources whose distances are greater than 3 kpc, their radial velocity is small enough with $|\vr|\le 15$ \kms, and galactic longitudes are either $0<l<80\deg$ or $280<l< 360\deg$. Also, sources in the Galactic Center direction at $340<l<20\deg$, which are mostly local objects, were omitted. Thus, we use here six sources as listed in table \ref{tab1}. Their positions in  the galactic plane are shown in figure \ref{plot} by filled circles.

Gwinn et al. (1992) measured the statistical distance to W49 to be $11.4 \pm 1.2$ kpc using VLBI observations of dispersions of radial velocities and proper-motions of maser sources. The source parameters are listed also in table \ref{tab1}, and the position is plotted in figure \ref{plot} by a filled square.  Roshi et al (2006) have obtained LSR velocities for associated molecular clouds of W49, which yields an average LSR velocity of $5.4 \pm 2.9$ \kms.

By applying the near-solar circle method to the data as listed in table \ref{tab1}, we calculated the galacto-centric distance $R_0$ by using equation (\ref{eq-rsun}), and list the results in table \ref{tab1}. The simply averaged value of the galacto-centric distance of the Sun is obtained  to be $R_0=7.54\pm 0.77$ kpc, whereas the weighted mean yields $R_0=7.13\pm 0.76$ kpc with individual weights proportional to $1/\delta R_0^2$. We here adopt the former as the result of the present analysis.

\def\skms{{\tiny \kms}}
\begin{table*}
\begin{center}
\caption{Distances and radial velocities of HII regions with  $|\vr|\le 15$ \kms, and derived $R_0$.}
\begin{tabular}{lllllllllll}
\hline\hline 
Source& $l$& $b$ & $r \pm \delta r$ & $v_r \pm \delta v_r$ & $R_0 \pm \delta R_0$ & Ref. to data\\
& (deg)& (deg)& (kpc)&  (\kms)   & (kpc)&   \\
\hline 
S104  &74.79 &$~~0.57$ & $4.40 \pm 1.40$ & $~~0.0 \pm 2.0$ & $8.39\pm 2.71$ &(1)\\
BBW287&283.76&$-3.41$& $3.73\pm 0.76$  &$-0.7 \pm 0.5$ & $8.05\pm 1.60$ &(1)\\
BBW324&287.00&$~~2.64$ & $3.13 \pm 0.49$ & $-13.5\pm 0.6$ &$8.11\pm 0.85$ &(1)\\
BBW311&287.22&$-3.05$& $3.10 \pm 0.63$ &$-7.7 \pm 0.5$ & $6.77\pm 1.07$ &(1)\\
BBW323&289.78&$-3.23$& $3.42\pm 0.70$  &$-14.3 \pm 0.5$ &$7.26\pm 1.04$ &(1)\\
BBW328&290.34&$-2.98$& $3.08\pm 0.37$  &$-12.9 \pm 0.5$ &$6.33\pm 0.54$ &(1)\\
W49 & 43.17 & $-0.10$& $11.40\pm 1.20$ &$ ~~5.4 \pm  2.9$ &$ 7.84\pm 0.75$ & (2), (3)\\
\hline 
Average &&&&& $7.54\pm 0.77$ \\
Weighted Avrg.&&&&& $7.13 \pm 0.76$\\
\hline
\end{tabular}
\\ 
\end{center} 
(1) Brand and Blitz (1993); (2) Gwinn et al. (1992); (3) Average of LSR velocities of associated clouds by Roshi et al. (2006)
\label{tab1}
\end{table*}

\begin{table*}
\begin{center}
\caption{VERA data and derived $R_0$ and $V_0$.}
\begin{tabular}{llllllllllllllll}
\hline\hline 
Source & $l$ & $b$ & $r \pm \delta r$ & $\vp \pm \delta \vp$  & $\vr \pm \delta \vr$ &$R_0 \pm \delta R_0$ & $V_0 \pm \delta V_0$ & Ref. to data\\
&(deg)&(deg)&(kpc) & (\kms)&(\kms) &  (kpc)&   (\kms) \\
\hline 
ON1&69.54&$-0.98$&$ 2.47 \pm 0.11 ^\dagger $& $-70.2 \pm  2.6$ & $12 \pm 1$ & $5.28 \pm 0.21$ & $154.5 \pm 5.8$ &(1) \\
ON2N&75.78&$-0.34$& $3.83\pm 0.13$ & $-104.6 \pm 2.9$  & $0 \pm 1$ & $7.80 \pm 0.39$ & $212.9 \pm 10.0$ &(2)\\


\hline 
\end{tabular}
\\ 
$\dagger$ Too close for the present method; (1) Nagayama et al. (2011); (2) Ando et al. (2011)\\
\end{center} 
\label{tab2} 
\end{table*}

\subsection{Determination of $R_0$ and $V_0$ from VERA data}
 
We try to apply the near-solar-circle method for simultaneous determination of  $R_0$ and $V_0$ to the star forming region ON1 using the recent VERA observations of H$_2$O maser lines as listed in table \ref{tab2} (Nagayama et al. 2011). . For reference, we also confirm the result for the solar circle object ON2N by  Ando et al. (2011). The velocities from VERA observations are referenced to the Local Standard of Rest after correction of the standard solar motion. Besides these sources, Ri et al. (2009a) compiled many VLBI trigonometric data, but the sources are too far from the solar circle, and are not used in this paper. 

Since the number of near solar circle objects is, thus, still limited, we only try to apply the method to ON1, and confirm the result for ON2N. The calculated results are shown in table \ref{tab2}. Since ON1 is too close to the Sun with $r=2.5$ kpc, it may be not appropriate for the present method. In figure \ref{plot} we plot the positions of the sources: the large triangle denotes ON2N, which is a solar-circle object, giving $R_0=7.8\pm 0.4$ kpc and $V_0=213\pm 10$ \kms (Ando et al. 2011), and the open triangle is ON1. 

\section{Discussion}

We have proposed a method to determine the galactic constants $R_0$ and $V_0$ from measurements of distances, radial velocities and proper motions of objects near the solar circle.
 
Determination of $R_0$ has been obtained for  HII regions listed in table \ref{tab1}. By averaging the derived values with equal weight, we obtain $R_0=7.54 \pm 0.77$ kpc. This value is consistent with the values of $R_0=7.6 \pm 0.3$ kpc (Eisenhauer et al. 2005) and $7.9 \pm 0.7$ kpc (Reid et al. 2009b)  measured for Sgr A$^*$ . It lies in the range from 6.7 to 8.9 kpc of the current values as obtained by the likelihood method by McMillan and Binney (2010). However, it is smaller than the value of $8.4\pm 0.6$ kpc derived by galactic model fitting by Reid et al. (2009a), although not inconsistent within the error. 

We have also tried simultaneous determination of $R_0$ and $V_0$ using proper motions from VERA observations for the two HII regions, ON1 and ON2N.  ON1 appears to be too close to the Sun ($r=2.5$ kpc) for application of the present method, resulting in unreasonable values, which may also be due to the source's intrinsic motion such as due to random motion and/or streaming motion. ON2N is an exceptional case, for which the strict solar circle method can be applied, as discussed by Ando et al (2011) in detail, and we have confirmed the values.

The present method is based on equations (\ref{eq-rsun}) and (\ref{eq-vsun}), simply applying Taylor expansion to the solar circle solutions by equations (\ref{eqr00}) and (\ref{eqv00}). Hence, the method does not require to assume any model for the rotation curve, except for the local $A$ value, which yields uncertainty on the second order of O($\vr \delta A/V_0 A$). This method may be, therefore, complimentary to the likely food analysis engaged by McMillan and Binney (2010), in which a model rotation curve is to be assumed.

Finally, we comment on possible intrinsic scatter in the kinematical observables that affect the results. Individual sources may have intrinsic velocity dispersion of $\delta \vr \sim$ several \kms on the same order of that for interstellar matter. Non-circular streaming motions would be also superposed on the circular motion. These will yield effective velocity dispersion on the order of $\delta \vr \sim \delta \vp \sim 10$ \kms, and result in the statistical error of the derived $R_0$ values on the order of $\delta R_0 \sim \pm 1$ kpc. To obtain a more precise $R_0$ value, we need a larger number of sources distributed over a wider area of the Galactic plane.

{}  

\end{document}